\documentclass{iaus}
\usepackage{graphicx}

\title[Maser Theory] %% give here short title %%
{Recent Developments in Maser Theory}

\author[Moshe Elitzur]   %% give here short author list %%
{Moshe Elitzur}

\affiliation{
Physics \& Astronomy Dept.,
University of Kentucky,
Lexington, KY 40506-0055, USA \break
email: moshe@pa.uky.edu
}

\pubyear{2007}
\volume{242}  %% insert here IAU Symposium No.
\pagerange{xxx--xxx}
\date{?? and in revised form ??}
\jname{ Astrophysical Masers and their Environments}
\editors{Jessica Chapman \& Willem Baan, eds.}

\def\eq#1{\begin{equation} {#1} \end{equation}}
\def\E#1{\hbox{$10^{#1}$}}
\def\about  {\hbox{$\sim$}}
\def\x      {\hbox{$\times$}}
\def\deg    {\hbox{$^\circ$}}
\def\kms    {\hbox{km\,s$^{-1}$}}
\def\ga     {\hbox{$\gtrsim$}}
\def\la     {\hbox{$\lesssim$}}
\def\mic    {\hbox{$\mu$m}}
\def\tV     {\hbox{$\tau_{\rm V}$}}
\def\LOH    {\hbox{$L_{\rm OH}$}}
\def\LIR    {\hbox{$L_{\rm IR}$}}
\def\Dv     {\hbox{$\Delta v$}}
\def\nuB      {\hbox{$\nu_B$}}
\def\DnuD     {\hbox{$\Delta\nu_D$}}
\def\xB       {\hbox{$x_B$}}           \let\xb=\xB
\def\Dm       {\hbox{$\Delta m$}}

\begin{document}

\maketitle

\begin{abstract}
This review covers selected developments in maser theory since the previous
meeting, {\em Cosmic Masers: From Proto-Stars to Black Holes} (Migenes \& Reid
2002). Topics included are time variability of fundamental constants, pumping
of OH megamasers and indicators for differentiating disks from bi-directional
outflows.


\end{abstract}

\firstsection

\section{Fun with Masers}

Rather then maser theory, the first couple of items covered here can be more
properly described as ``fun with masers''. These items were selected for
discussion because they do not belong under the title of any other talk in this
meeting. Both involve weak OH signals that lead to spectacular results.

\subsection{How do we know masers exist?}

There are numerous arguments why the radiation from various astronomical
sources involves maser amplification (e.g., Elitzur 1992). But the only direct
proof can come from measurements of excitation temperatures that yield negative
values. The well known technique for measuring a line excitation temperature,
$T_x$, requires a molecular cloud in front of a distance continuum source that
provides background radiation with measurable brightness temperature $T_b$ at
the transition frequency. The measured brightness temperatures in the ``off''
and ``on'' positions, away from and toward the background source, obey
\[
    T^{\rm off} = T_x\left(1 - e^{-\tau}\right), \qquad
    T^{\rm on}  = \left(T_x - T_e\right)\left(1 - e^{-\tau}\right)
\]
where $\tau$ is the line optical depth of the cloud. The two measurements yield
two equations for the two unknowns $T_x$ and $\tau$. In the classic experiment
by Rieu \etal\ (1976), the OH main lines from a cloud in front of the
extragalactic radio source 3C123 switched from emission in the off-source
position to on-source absorption. The excitation temperature of both lines is
$T_x \simeq$ 5--7 K. But in the case of the 1720 MHz line, off-source emission
turned into even stronger emission in the on-source position, indicating that
the cloud amplifies the background radiation with $\tau \simeq -0.1$. The
corresponding excitation temperature is $T_x(1720) \simeq -10$~K.

Almost 30 years later, Weisberg \etal\ (2005) came up with a beautiful variant
of this approach in OH spectral measurements toward 18 pulsars.  One member of
the sample, pulsar B1641--45, emits every 0.455 seconds a pulse of radio
radiation that passes through an OH cloud on its way toward Earth. The entire
system is positioned in front of an H{\small II} region that provides
additional, steady background radiation. When the pulsar is off, spectra of the
OH ground-state lines detected from the cloud display absorption features in
the 1612, 1665 and 1667 MHz lines and an emission feature at 1720 MHz. When the
pulsar is on, passage of its radiation through the cloud deepens the absorption
features. But at 1720 MHz the emission feature becomes stronger, showing that
at this frequency the pulsar radiation is amplified by passing through the
intervening screen. This is the equivalent of an object amplifying a background
light instead of casting a shadow. It is the same effect as observed earlier by
Rieu \etal, but it does not involve any change in the telescope pointing. The
pulsar method also increases confidence in the results, thanks to the repeated
detection of amplified signals that arrive more than twice every second. The
amplification again is modest, $\tau$ is only $\sim -0.05$. Although the input
signals are amplified by only a few percent in both cases, these weak masers
are the only unambiguous direct evidence for amplification.

%%%%%%%%%%%%%%%%%% Figure %%%%%%%%%%%%%%%%%%%%%%%%%%%%%%%%%%%%%%%

\begin{figure*}[ht]
  \centering
  \begin{tabular}{lr}
    \begin{minipage}{0.40\hsize}

\caption{ OH satellite line spectra toward PMN J0134–0931 plotted against
barycentric velocity relative to the source redshift, $z$ = 0.76355. (A) 1720
MHz transition; (B) 1612 MHz transition; (C) sum of the 1612 and 1720 MHz
spectra; this spectrum is consistent with noise. (Kanekar \etal\
2005)}\label{fig:Kanekar}
   \end{minipage}
    &
    \begin{minipage}{0.5\hsize}
      \centering
      \includegraphics[width=0.9\hsize,clip]{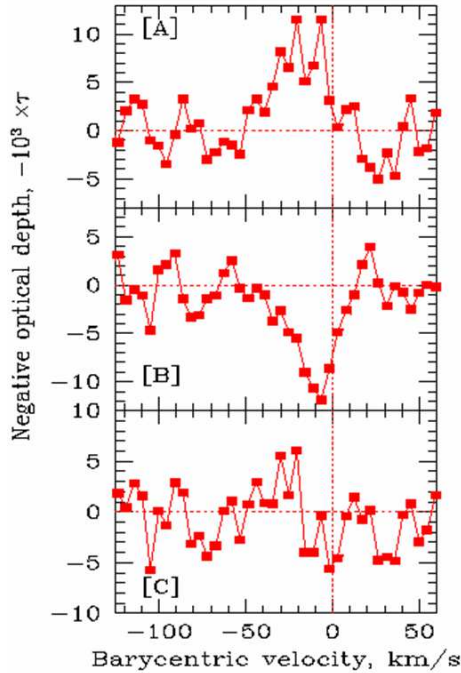}
      \end{minipage}
  \end{tabular}
\end{figure*}

%%%%%%%%%%%%%%%%%%%%%%%%%%%%%%%%%%%%%%%%%%%%%%%%%%%%%%%%%%%%%%%%%

\subsection{Are fundamental constants truly constant?}

A cosmological implication of many variants of string theories is that
fundamental constants such as the fine structure constant $\alpha = e^2/\hbar
c$, the electron-proton mass ratio $m_{\rm e}/m_{\rm p}$, etc.\ should vary
with time. Indeed, from measurements of QSO absorption lines it has been
suggested that a variation $\Delta\alpha/\alpha$ \about\ \E{-5} has possibly
occurred between the $z$ = 3.5 epoch and the present (Murphy \etal\ 2001).
However, the uncertainty in this suggestion is substantial because measurements
based on atomic lines are faced with fundamental difficulties. Comparing the
frequencies of different lines of the same atom does not yield information on
$\alpha$ because the entire spectrum of a given species has the same
$\alpha$-dependence, therefore frequency shifts that would result from a change
in $\alpha$ cannot be distinguished from the effects of redshift. And
comparison of lines of different atoms suffers from the uncertainty arising
from the possibly different locations, and different $z$, of the species.

Darling (2003) pointed out that the complexity of the OH spectrum makes it most
suitable for getting around this problem. The centimeter OH transitions are a
combination of hyperfine splitting and $\Lambda$-doubling, and the two effects
depend differently on $\alpha$. As a result, by comparing the frequencies of
different OH radio transitions one can constrain the cosmic evolution of
$\alpha$ from a single species, avoiding systematic errors in $\alpha$
measurements from multiple species which may have relative velocity offsets.
The problem of different locations still exists, though, because different OH
lines may still arise from different locations because of preferential
excitations in different environments. This problem was solved by Kanekar
\etal\ (2005) who noted that many weak signals from the OH ground-state
satellite lines appear in conjugate pairs of absorption and emission. The two
halves of the $\Lambda$-doublet consist of pairs of hyperfine states. Because
of the different overall spins of hyperfine states, their coupling to different
excited states makes it easy to transfer population between them without
affecting the overall population of each half of the $\Lambda$-doublet. The
result is an overpopulation, or even inversion, of one satellite line and a
mirror under-population of the other (see, e.g., Elitzur 1992).

Taking OH spectra from the $z$ \about\ 0.765 gravitational lens toward PMN
J0134–0931, Kanekar \etal\ identified a conjugate pair of 1612 MHz absorption
and 1720 MHz emission, shown in Figure \ref{fig:Kanekar}. As is evident from
the figure's bottom panel, the sum of the two satellite lines produces a flat
spectrum consistent with noise, a strong indication that they originate from
the same region. The deduced line frequencies place tight bounds on the
variation of fundamental constants over the past \about\ 6.5 G yr:
$\Delta\alpha/\alpha < $ 6.7\x\E{-6} and a similarly strong upper limit of
1.4\x\E{-5} for $m_{\rm e}/m_{\rm p}$. These fundamental constants have not
changed, to within this tight tolerance, for half the age of the Universe.

\section{OH Megamasers}

OH megamasers (OHM) are extremely luminous extragalactic sources with isotropic
luminosities a million or more times greater than their Galactic OH maser
counterparts. They are found in the nuclear regions of luminous infrared
galaxies where intense star formation is occurring. The ``standard model" for
OHM was introduced by Baan (1985), who proposed that the maser emission is
produced by low gain unsaturated amplification of background radio continuum.
This proposal was confirmed by the Henkel \& Wilson (1990) comprehensive study
of OHM. However, compact maser emission discovered in subsequent VLBI
observations produced conflicts with the standard model, conflicts that reached
their climax around the time of the previous maser meeting (Lonsdale 2002).
Strong compact masers without correspondingly bright continuum background
implied amplification factors in excess of 800. In addition, the 1667/1665 line
ratios of the compact masers greatly exceeded those of the diffuse OHM
emission. An additional surprise of these observations was that linewidths
remained large (tens of \kms) even on the smallest observed angular scales.
This has led to the suggestion that there are two different modes of OHM
operation --- low gain, unsaturated amplification responsible for the diffuse
maser emission and high gain, saturated emission producing the compact sources.
It was also suggested that the two classes may have different pumping
mechanisms (Lonsdale 2002).

A major recent development is the realization that a two-component picture is
not necessary. The compact masers can be explained within the standard model as
chance alignment of low-gain maser clumps. The key was the interferometric
observations of IIIZw35 by Pihlstr{\"o}m \etal\ (2001) and the modeling by the
same authors and later, in greater detail, by Parra \etal\ (2005). The model
explains both the diffuse and compact emission from IIIZw35 by assuming a
clumpy maser medium. Each cloud generates low gain, unsaturated emission, and
strong compact emission occurs when the line of sight intersects more than one
cloud. A similar model has subsequently been used to explain the megamaser
emission from IRAS 17208-0014 (Momjian \etal\ 2006). The clumpy maser model
also explains the observation that compact masers are always found embedded in
the diffuse emission and do not occur in isolation (Lonsdale 2002). Thus the
only additional ingredient necessary to supplement Baan's ``standard model'' is
clumpiness --- the OHM phenomenon is explained as the amplification of
background radio continuum by low gain ($\tau_{1667}$ \la\ 1) clumps.

Lockett \& Elitzur (2007) performed pumping calculations to explain the OHM
inversion process. Pumping models should explain the outstanding properties of
OHM emission: (1) The line ratio of observed fluxes always obeys
$F(1667)/F(1665) > 1$ and the satellite lines are always weaker than the main
lines. By comparison, in evolved stars the main lines emission region displays
a similar behavior --- 1667 MHz is the strongest maser, and the satellite line
emission from that region is weaker than either main line. In contrast,
Galactic star forming regions show a different, more complex emission pattern.
Generally, the main lines are stronger than the satellite lines and display
$F(1667)/F(1665) < 1$, the opposite of EHM and evolved stars. But there are
sources in which the strongest line is the 1665 MHz is and some where it is a
satellite line.

%%%%%%%%%%%%%%%%%% Figure %%%%%%%%%%%%%%%%%%%%%%%%%%%%%%%%%%%%%%%

\begin{figure}[ht]
 \centering \leavevmode
 \includegraphics[height=0.4\hsize,clip]{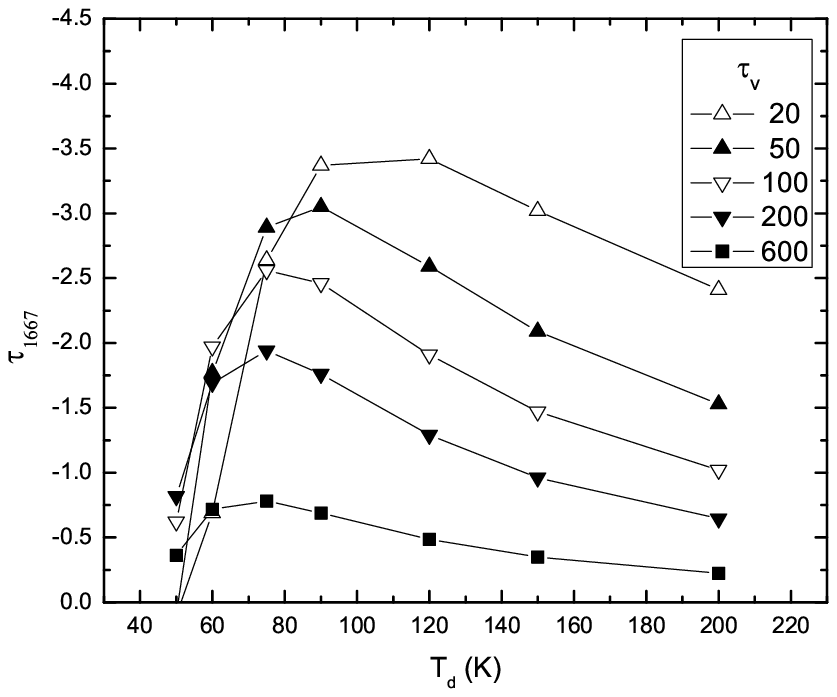} \hfill
 \includegraphics[height=0.4\hsize,clip]{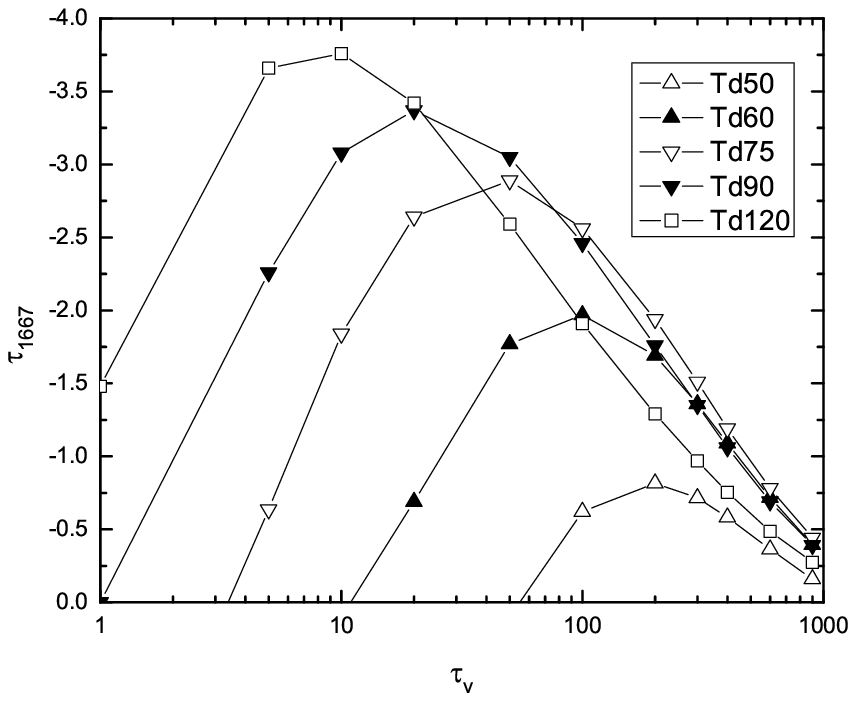}

\caption{ Dependence of the 1667 MHz maser optical depth, $\tau_{1667}$, on the
parameters of the IR pumping flux (see eq.\ \ref{eq:Dust}). {\em Left:}
$\tau_{1667}$ versus dust temperature for varying dust optical depth. {\em
Right:} $\tau_{1667}$ versus dust opacity for different dust temperatures
(Lockett \& Elitzur 2007).
}\label{fig:FIR}
\end{figure}

%%%%%%%%%%%%%%%%%%%%%%%%%%%%%%%%%%%%%%%%%%%%%%%%%%%%%%%%%%%%%%%%%

In the Lockett \& Elitzur models, the inversion arises from radiative
excitations by far-IR radiation (FIR). The FIR radiation field is described by
the simple expression
\eq{\label{eq:Dust}
      F_\nu = B_\nu(T_d)\left(1 - e^{-\tau_\nu}\right),
}
corresponding to emission from dust with a uniform temperature $T_d$ and
optical depth $\tau_\nu$. The frequency dependence of $\tau_\nu$ follows that
of standard interstellar dust so the only free parameters are $T_d$ and
$\tau_V$, the dust optical depth at visual. Combinations of parameters that
produce main lines inversion are shown in figure \ref{fig:FIR}. The left panel
shows the 1667 MHz optical depth versus dust temperature for a range of dust
opacities. A minimum dust temperature of \about\ 45 K is needed for maser
production. Increasing the dust temperature increases the maser strength up to
a point, after which it decreases. Maximum maser inversion occurs for dust
temperatures between \about\ 70 -- 120 K. The right panel shows the 1667 MHz
optical depth versus dust opacity for a range of dust temperatures. Two
important results can be seen in this figure. First, when the opacity becomes
large the dust radiation approaches that of a black-body and the maser begins
to thermalize. At \tV\ \ga\ 200 there is little dependence of maser strength on
dust temperature. Second, for each dust temperature there is a minimum dust
opacity needed for inversion. This implies that a minimum FIR pumping flux is
needed for maser production. Further understanding can be found by examining
the maser dependence on the FIR intensity in the maser pump line at 53 \mic.
The pumping flux is determined by the product of the two factors in equation
\ref{eq:Dust}. The first is the Planck function, which depends on dust
temperature, the second is the self-absorption term, which depends on the dust
opacity at the pumping frequency. The FIR flux is increased by increasing
either the dust temperature or its opacity. As a result, a larger flux is
needed for a lower dust temperature, indicating that the FIR spectral shape is
important for maser pumping. Thus a smaller IR luminosity can support inversion
as long as the dust temperature is sufficiently high --- in agreement with
observations (Baan 1989, Darling \& Giovanelli 2002). Similarly,  inversion
requires a minimum dust temperature. Below \about\ 50 K, the Planck function at
53 \mic\ decreases very rapidly with dust temperature. The flux can be
increased by raising the dust opacity but cannot exceed the black-body limit,
which is insufficient for inversion at dust temperatures below about 45 K.

An important correlation is the increase of \LOH\ with \LIR. Early OHM
observations revealed an approximately quadratic relation between the two (Baan
1989), but subsequent studies established the flatter dependence $\LOH\
\varpropto \LIR^{1.2}$ (Darling \& Giovanelli 2002). Low gain amplification of
background radio continuum implies that \LOH\ is proportional to the radio
luminosity, which in turn is proportional to \LIR\ over the relevant luminosity
range. Since the observed correlation is steeper than linear, Baan (1989)
conjectured that maser optical depth $\tau_{\rm maser}$ also varies linearly
with the pump luminosity \LIR. However, the Lockett \& Eitzur model results
show that, depending on the different parameters, $\tau_{\rm maser}$ in fact
can either increase or decrease with the IR luminosity. The increase of \LOH\
in excess of linear proportion to \LIR\ may be explained by the results of the
Gao \& Solomon (2004) of HCN emission from luminous IR galaxies. They find a
tight proportionality between IR and HCN luminosities, and conclude that the
amount of dense molecular gas, which is traced by the HCN, is linearly
proportional to \LIR. It is thus reasonable to assume that, together with the
increase of dense molecular gas, the number of OH maser clouds is also
increasing with \LIR.

%%%%%%%%%%%%%%%%%% Figure %%%%%%%%%%%%%%%%%%%%%%%%%%%%%%%%%%%%%%%

\begin{figure}
 \centering \leavevmode
 \includegraphics[width=0.6\hsize,clip]{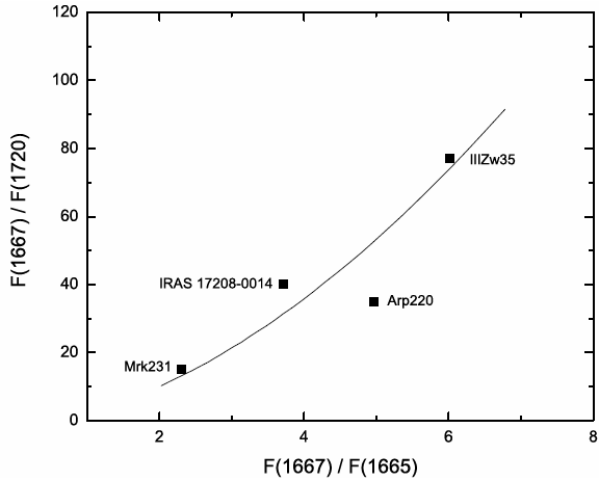}

\caption{``Color-color" diagram of OH maser line ratios. The line shows the
variation when all four lines have the same excitation temperature, which is
the situation when the linewidth exceeds \about\ 10 \kms, as $\tau_{1667}$
increases from 0.5 at the curve's lower end to 4 at its upper end. Data are
from Baan \etal\ 1989, 1992 and Martin et al 1989 (Lockett \& Elitzur 2007).
}\label{fig:ratios}
\end{figure}

%%%%%%%%%%%%%%%%%%%%%%%%%%%%%%%%%%%%%%%%%%%%%%%%%%%%%%%%%%%%%%%%%

Line overlap is a major component of the OHM inversion and the calculations
show that the linewidth, \Dv, is the most important factor in determining the
1667/1665 MHz line ratio. Widths in excess of \about\ 2 \kms\ cause the
1667/1665 ratio to be consistently larger than one, while widths \la\ 2 \kms\
can cause this ratio to be greater than or less than one, depending on the
actual \Dv. Large linewidths (\ga\ 10 \kms) produce ground state hyperfine
transitions that are all inverted with approximately the same (negative)
excitation temperature. Detailed comparison with observations is difficult
because the satellite lines are very weak and have only been detected for
relatively nearby sources. Even in those cases the signal to noise ratio is
small, resulting in large uncertainties in line strength ratios. Figure
\ref{fig:ratios} plots the 1667/1720 MHz line ratio versus the 1667/1665 MHz
line ratio for four relatively nearby OHM. The agreement is surprisingly good
considering the large uncertainties in the line ratios.

The variation of line ratios with \Dv\ suggests that FIR might be a common pump
of all OH main line masers, Galactic as well as megamasers, with the magnitude
of velocity controlling line overlap the factor explaining the observed
differences. The maser patterns in Galactic star forming regions stand apart
because of the relatively small values of \Dv\ in these sources. The similarity
between OHM and evolved stars may reflect the large velocities controlling line
overlap in both environments.

%%%%%%%%%%%%%%%%%% Figure %%%%%%%%%%%%%%%%%%%%%%%%%%%%%%%%%%%%%%%

\begin{figure*}[ht]
  \centering
  \begin{tabular}{cc}
    \begin{minipage}{0.50\hsize}
 \caption{%\small
Position-velocity diagram of 12 GHz methanol maser amplification
contours in NGC 7538 IRS 1N (Pestalozzi \etal\ 2004).}
\label{fig:NGC7538}
    \end{minipage}
    &
    \begin{minipage}{0.45\hsize}
      \centering
 \includegraphics[bb=53 195 284 380, width=0.9\hsize,clip]{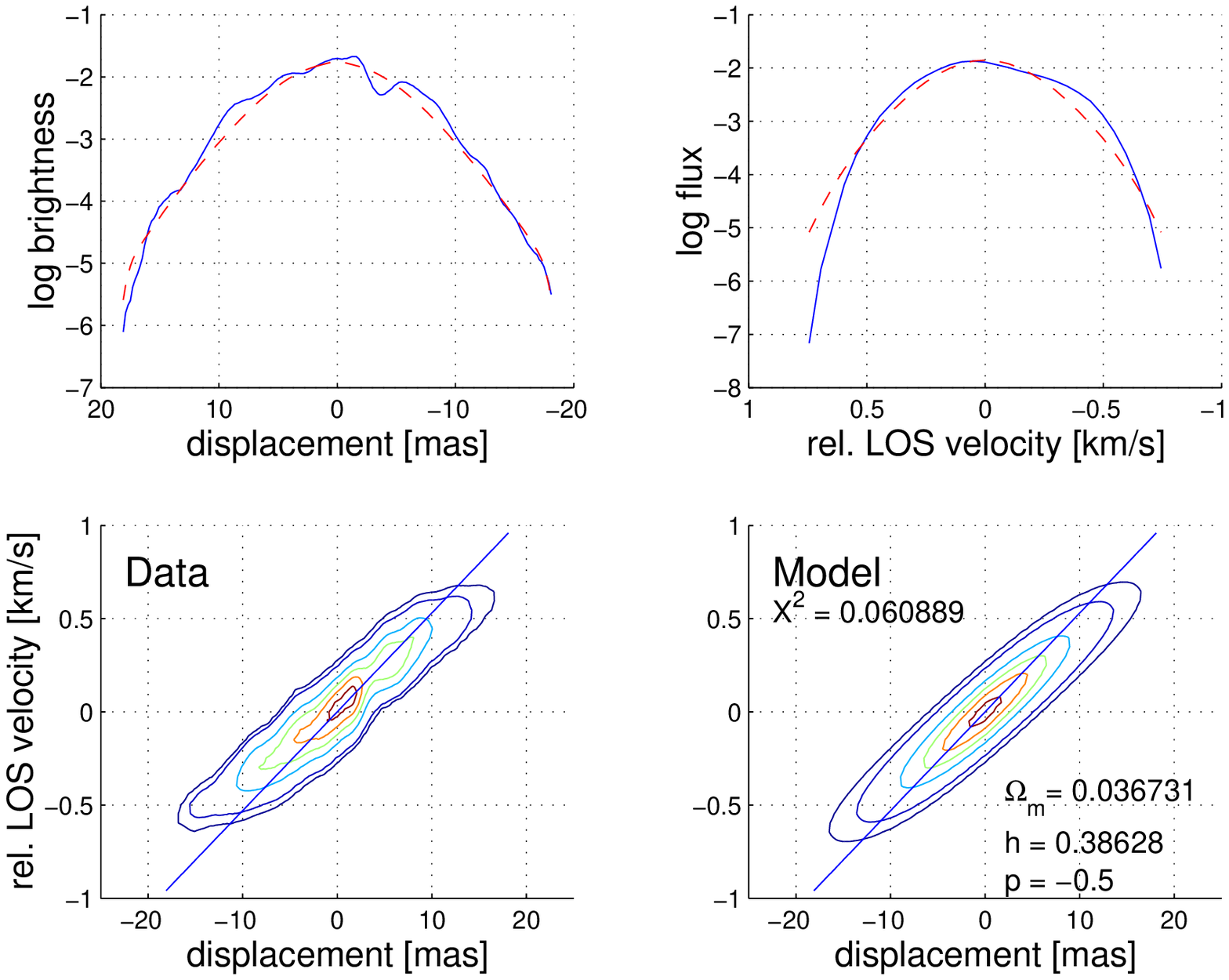}
      \end{minipage}
  \end{tabular}
\end{figure*}

%%%%%%%%%%%%%%%%%%%%%%%%%%%%%%%%%%%%%%%%%%%%%%%%%%%%%%%%%%%%%%%%%

\section{Disks and Outflows}

Class II methanol maser emission often shows linear structures in both spatial
maps and in position--velocity ($p$--$v$) diagrams. It has been suggested that
these structures delineate protostellar edge-on disks (Norris \etal\ 1998).
However, similar structures can be produced also in bi-directional outflows, a
geometry that actually seems more appropriate in a large fraction of methanol
masers (De Buizer 2003). The fact that such disparate maser geometries can
produce such similar signatures raises a fundamental question: Short of proper
motion measurements, is there a reliable way to distinguish between the two?
The methanol maser source toward IRS 1N in NGC 7538 may offer an answer. Its
unique emission contours display a centrally-peaked large, elongated structure
in the $p$--$v$ diagram, shown in figure \ref{fig:NGC7538}; in contrast, most
other sources typically display in such diagrams a string of maser features.
The structure in NGC 7538 was successfully modeled in great detail by
Pestalozzi \etal\ (2004) with an edge-on Keplerian disk. Although this
interpretation was questioned by De Buizer \& Minier (2005), Kraus \etal\
(2006) have demonstrated the existence of jet precession in this source,
settling the issue in favor of the disk geometry.

%%%%%%%%%%%%%%%%%% Figure %%%%%%%%%%%%%%%%%%%%%%%%%%%%%%%%%%%%%%%

\begin{figure}
 \centering \leavevmode
 \includegraphics[width=\hsize,clip]{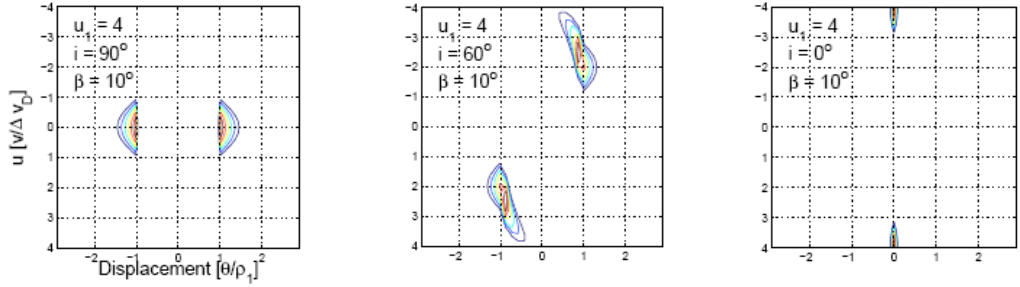}

\caption{Position-velocity diagram of maser amplification contours for bipolar
outflow with representative properties. The outflow axis is in the plane of the
sky in the left panel, aligned with the line-of-sight in the right panel and
60\deg\ off the line-of-sight in the center one (Pestalozzi \etal\ 2007).
}\label{fig:outflow}
\end{figure}

%%%%%%%%%%%%%%%%%%%%%%%%%%%%%%%%%%%%%%%%%%%%%%%%%%%%%%%%%%%%%%%%%

%%%%%%%%%%%%%%%%%% Figure %%%%%%%%%%%%%%%%%%%%%%%%%%%%%%%%%%%%%%%

\begin{figure}
 \centering \leavevmode
 \includegraphics[width=0.45\hsize,clip]{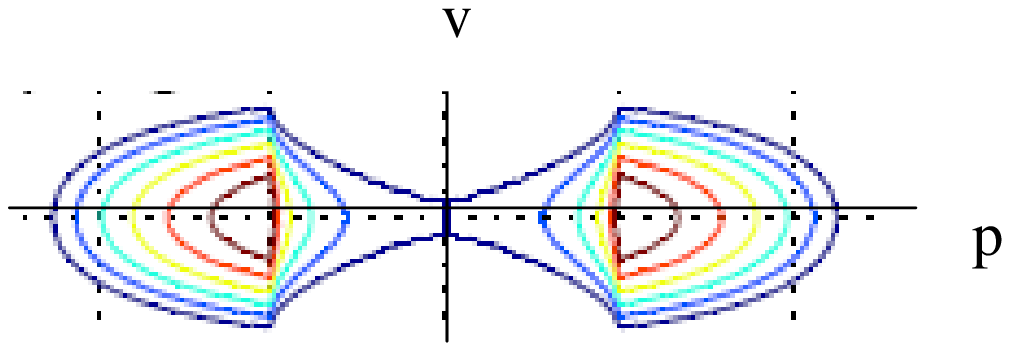} \hfill
 \includegraphics[width=0.45\hsize,clip]{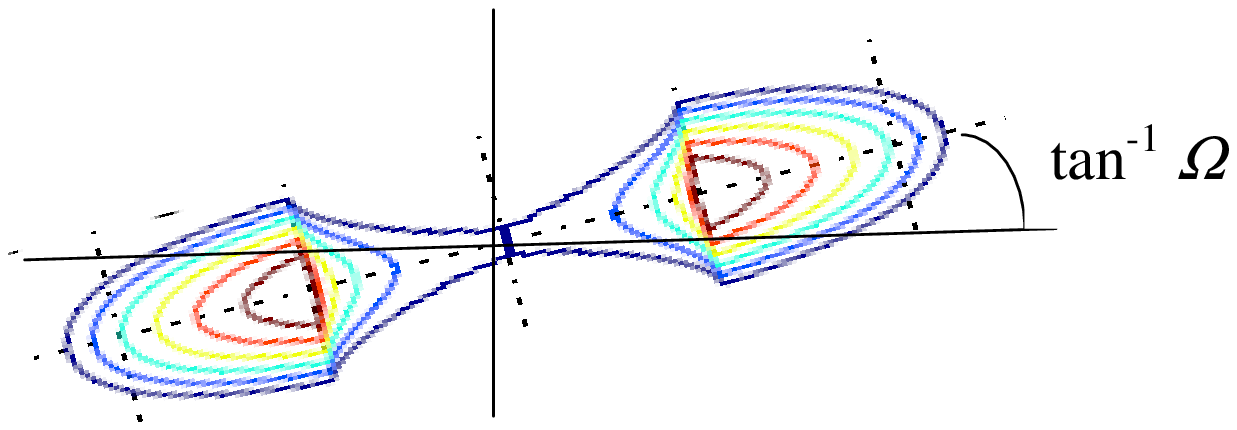}

\caption{Position-velocity diagram for maser amplification in an edge-on disk.
{\em Left}: A quiescent disk, only thermal velocities. {\em Right}: The same
disk subjected to solid-body rotation with angular velocity $\Omega$
(Pestalozzi \etal\ 2007).
}\label{fig:disk}
\end{figure}

%%%%%%%%%%%%%%%%%%%%%%%%%%%%%%%%%%%%%%%%%%%%%%%%%%%%%%%%%%%%%%%%%

It turns out that the $p$--$v$ structure seen in figure \ref{fig:NGC7538}
implies fairly tight constraints that significantly narrow the range of
possible geometries (Pestalozzi \etal\ 2007). Consider first a bipolar outflow.
When the outflow axis is in the plane of the sky, the emission from each lobe
is centered on velocity $v = 0$. Because the density declines rapidly with
radial distance, the emission is peaking at two regions displaced symmetrically
from the outflow origin. The resulting structure in the $p$--$v$ diagram is two
blobs that straddle the origin at $v = 0$, as seen in the left panel of figure
\ref{fig:outflow}. When the outflow axis is along the line-of-sight, the
emission from both cones occurs at the same position, but at different
velocities --- blue and red shifted by the same amount from the systemic
velocity. The $p$--$v$ diagram, shown in the right panel of figure
\ref{fig:outflow}, again displays two blobs, only this time they are in the
same position but at different velocities. In essence, rotating the outflow
axis by 90\deg\ effects a similar rotation of the emission contours in the
$p$--$v$ diagram. It is not surprising that an intermediate orientation of the
outflow axis likewise produces an intermediate rotation in the $p$--$v$
diagram, as can be seen from the center panel of figure \ref{fig:outflow}.
Whatever the inclination of its axis, a bipolar outflow cannot produce the
single centrally peaked structure seen in figure \ref{fig:NGC7538}. A smooth
outflow will produce two mirror features displaced symmetrically from the
center. Fragmentation of the outflow may introduce an asymmetry between the two
halves but will always avoid the center, never producing a centrally peaked
feature.

Consider now the case of an edge-on disk. When the disk does not rotate, its
optical depth obeys $\tau(p,v) = \tau(p)\phi(v)$; that is, the positional
variation of the amplification is controlled solely by the length of the
amplifying column because the velocity profile $\phi$ is the same at every
position. For uniform inversion, $\tau(p)$ is simply proportional to the
pathlength through the disk, which is maximal at the two tangents to the disk
inner radius. The $p$--$v$ diagram of the amplification contours in this case
is shown in the left panel of figure \ref{fig:disk}. Similar to the outflow
geometry, the contours show two distinct peaks, displaced symmetrically from
the center. The center does show amplification in this case, but it is a local
minimum. When the disk is set in rotation, the Doppler velocity of the emission
from a point at radius $r$ and impact parameter $p$ obeys $v = \Omega(r)p$,
where $\Omega(r)$ is the local angular velocity. Therefore, in solid body
rotation the material along each impact parameter remains fully velocity
coherent, only the velocity profile is centered on $v = \Omega p$; that is, the
optical depth obeys $\tau(p,v) = \tau(p)\phi(v - \Omega p)$. As a result, the
structure of amplification contours in the $p$--$v$ diagram remains the same,
only rotated by the angle tan$^{-1}\Omega$ (and slightly stretched to maintain
the peak positions at the two tangents to the inner radius), as shown in the
right panel of figure \ref{fig:disk}. Similar to bipolar outflow, an edge-on
disk rotating as a solid body will never produce a centrally peaked structure
as observed in NGC 7538 (figure \ref{fig:NGC7538}). Only differential rotation,
such as Keplerian motion, can produce such structure. Furthermore, even in that
case, turning the local minimum at the center of the $p$--$v$ diagram into a
local maximum requires rotation rates in excess of a certain threshold
(Pestalozzi \etal\ 2007). Detailed analysis of the $p$--$v$ diagram is,
evidently, a powerful tool that deserves more study.

\section{Open Problem --- Maser Polarization}

Maser polarization behaves differently according to whether the Zeeman shift
\nuB\ is larger or smaller than the linewidth \DnuD. The case $\nuB \gg \DnuD$
is well understood. Spontaneous decays occur in pure \Dm\ transitions,
producing in each transition radiation that is fully polarized and centered on
a different Zeeman frequency. Under these circumstances the amplification
process preserves the polarization, therefore thermal and maser polarizations
are the same. The only difference between the two cases is the disparity
between the $\pi$ and $\sigma$ maser intensities, reflecting their different
growth rates (Elitzur 1996).

In the opposite limit, $\nuB \gg \DnuD$, the Zeeman components overlap and the
amplification mixes the different polarizations. The pumping processes produce
radiation in three different modes of polarization with respect to the magnetic
field ($\Delta m = 0, \pm1$), but only two independent combinations propagate
in any direction because the electric field must be perpendicular to the wave
vector. Elimination of the longitudinal component implies specific phase
relations among the three pump-generated electric fields; only waves launched
with these phase relations produce superpositions that are purely transverse so
that they can be amplified by propagation in the inverted medium. The phase
relations translate to the following polarization for maser radiation
propagating at angle $\theta$ from the magnetic axis:
\eq{
  {Q\over I} = -1 + {2\over3\sin^2\theta},  \qquad
  {V\over I} = {16x\xb\over3\cos\theta}\,,
}
where $x = (\nu - \nu_0)/\DnuD$ and \xb\ = \nuB/\DnuD. This solution, which was
first derived by Goldreich, Keeley \& Kwan (1973) in the limit \xB\ = 0 and
extended by Elitzur (1996) to finite $\xB < 1$, is obtained also from the
requirement that the four Stokes parameters produce fractional polarizations
that remain unaffected by the amplification process.

How the unpolarized radiation produced in spontaneous decays evolves into this
stationary polarization solution remains an open problem. Numerical
calculations that follow the growth of the Stokes parameters do not reach the
stationary solution for any reasonable amount of amplification, and this slow
growth can be understood from simple analytic considerations. However, the
Stokes parameters are intensities, quadratic in the electric fields of the
waves, and in formulating their growth equations all information about the
phases of the electric vectors is lost. Since the stationary solution reflects
specific phase relations between the $\pi$ and $\sigma$ components, the
formalism employed for calculating the growth of the Stokes parameters negates
the very effect it seeks to describe. Eliminating the phase information, the
interactions selecting the stationary polarization are removed from the
calculation. What is still missing is a full simulation of the ensemble
evolution of interacting particles and electric fields. Such simulations are
preformed in studies of laboratory plasma but have not yet been attempted for
astronomical maser radiation.

\section{Coupled Escape Probability (CEP)}

Pumping models require reliable calculations of the population distribution in
multi-level systems. The radiative terms in the statistical rate equations
require the local radiation field, which must be determined from a separate
solution of the radiative transfer equation for each and every line. Because of
the complexity and computational demands of exact solution methods, most codes
bypass altogether the radiative transfer equations, employing instead the
escape probability approximation. In this approach only the level populations
are considered, calculated from rate equations that include photon escape
factors which are meant to account approximately for the effects of radiative
transfer. This approach amounts to an uncontrolled approximation without
internal error estimates because it is founded on a plausibility assumption
right from the start. Nevertheless, this inherent shortcoming is often
tolerated because of the simplicity and usefulness of the escape probability
approach.

Elitzur \& Asensio Ramos (2006) have recently developed a new exact method, the
Coupled Escape Probability (CEP), that retains all the advantages of the naive
escape probability approach. In this new technique the source is divided into
zones, and formal level population equations are derived rigorously from first
principles. Different zones are coupled through terms resembling standard
escape probability expressions, resulting in a set of coupled algebraic
equations for level populations that are self-consistent with the line
radiation they generate. Any desired accuracy can be achieved by increasing the
number of zones. In comparative studies of a number of standard problems, the
CEP method outperformed the current leading techniques by substantial margins.
While the new method holds a great performance edge, its greatest advantage is
its simplicity and ease of implementation. The CEP method employs a set of
algebraic equations that are already incorporated in all codes based on the
escape probability approximation. All that is required for an exact solution is
to augment the escape probability in such codes with a zone-coupling sum. With
this simple modification, the multi-level line transfer problem is solved
exactly and efficiently.

For my own pumping calculations I developed over the years the escape
probability code MOLPOP that can handle any arbitrary atom or molecule for
which atomic data exist. For any species, the energy levels, collision rates
and Einstein A-coefficients are tabulated in ordinary ASCII text files and
MOLPOP selects from its database the appropriate files according to input
instructions. We are in the process of implementing the CEP technique into
MOLPOP, creating a general purpose tool for atomic and molecular line analysis
that will become publicly available. In addition, the atomic and molecular data
team at Meudon Observatory has embarked on the development of another tool that
accesses through the web their central database and creates on the user's local
computer atomic and molecular data files in the format required by MOLPOP. This
auxiliary tool will become part of the MOLPOP/CEP package and will provide
users the most current atomic data at all times. The complete package will
enable exact pump modeling for every maser molecule. We hope to have a first
release by the end of 2007.

\begin{acknowledgments}
The author's maser research is supported by NSF through grant AST-0507421.
\end{acknowledgments}

\end{document}